\documentclass[preprint,preprintnumbers,showpacs,amsmath,amssymb,nofootinbib]{revtex4}
\usepackage{graphicx}% Include figure files
\usepackage{dcolumn}% Align table columns on decimal point
\usepackage{bm}% bold math
\begin{document}
\title{On the test of the modified BCS at finite 
temperature}
\author{Nguyen Dinh Dang$^{1}$ and Akito Arima$^{2}$}
\affiliation{%
$^{1}$ Cyclotron Center, RIKEN, 2-1 Hirosawa, Wako, 351-0198 
 Saitama, Japan\\
$^{2}$ Japan Science Foundation, Kitanomaru-Koen, Chiyoda-ku, 
102-0091 Tokyo, Japan
}%

%\date{\today}% It is always \today, today,
	     %  but any date may be explicitly specied
\begin{abstract}
    The results and conclusions by Ponomarev and Vdovin [Phys. Rev. C {\bf 72}, 
034309 (2005)]
    are inadequate to judge the applicability of the modified BCS
    because they were obtained either in the temperature region, 
    where the use of zero-temperature single-particle spectra is no 
    longer justified, or in too limited configuration spaces.  
\end{abstract}
\maketitle
The modified BCS theory (MBCS) was proposed and developed in
\cite{MBCS1,MBCS2,MHFB} as a microscopic approach to take into 
account fluctuations of quasiparticle numbers, which the BCS theory
neglects. The use of the MBCS in nuclei at finite 
temperature $T$ washes out the sharp superfluid-normal phase 
transition. This agrees with the predictions 
by the macroscopic theory~\cite{Moretto}, the exact 
solutions~\cite{Egido}, and  experimental data~\cite{exp}.
The authors of \cite{test} claimed that the MBCS is 
thermodynamically inconsistent and its applicability is far below the 
temperature where the conventional BCS gap collapses.
The present Comment points out 
the shortcomings of \cite{test}.  We concentrate only on the major issues 
without repeating minor arguments already discussed in 
\cite{MBCS2,MHFB}
or inconsistent comparisons in Fig. 9 and footnote [11] of 
\cite{test} (See ~\cite{note}).

1) The application of the statistical formalism in finite nuclei requires 
that $T$ should be small 
compared to the major-shell spacings ($\sim$ 5 MeV for 
$^{120}$Sn). In this case 
zero-$T$ single-particle energies can be extended to $T\neq$ 0. 
As a matter of fact, the $T$-dependent Hartree-Fock (HF) calculations for heavy nuclei in \cite{Bonche}
have shown that already at $T\geq$ 4 MeV the effect of $T$ on 
single-particle energies cannot be neglected.
We carried out a test 
calculation of the neutron pairing gap for $^{120}$Sn, 
where, to qualitatively mimic the compression of the 
single-particle spectrum at high $T$ as in \cite{Bonche}, 
the neutron energies are $\epsilon_{j}'=\epsilon_{j}(1+\gamma T^{2})$ with
$\gamma=-1.2\times 10^{-4}$ if $|j\rangle\leq |1g_{9/2}\rangle$. For  
$|j\rangle$ above $|1g_{9/2}\rangle$, we took $\gamma$ equal to 
$0.49\times 10^{-3}$ and $-0.7\times 10^{-3}$ for negative and 
positive $\epsilon_{j}$, respectively. The obtained MBCS gap has a smooth and 
positive $T$ dependence similar to the solid line in Fig. 7 of 
\cite{MBCS1} with a flat tail of around 0.2 MeV from $T=$ 5 MeV up to $T=$ 7 MeV. 
For the limited spectrum used in the calculations of Ni 
isotopes~\cite{MBCS2}, 
the major-shell spacing between (28-50) and (50-82) shells 
is about 3.6 MeV, so the region of valid 
temperature is $T\ll$ 3.6 MeV. 
Hence, the strange behaviors in the results obtained 
at large $T$ for $^{120}$Sn and
Ni isotopes in \cite{test} occurred because the zero-$T$ 
spectra were extended to too high $T$. Moreover, the 
configuration spaces used for Ni isotopes are too small for the 
MBCS to be applied at large $T$. The same situation takes place 
within the picket-fence model (PFM) analyzed below.

\begin{figure}                                                             
\includegraphics[width=20pc]{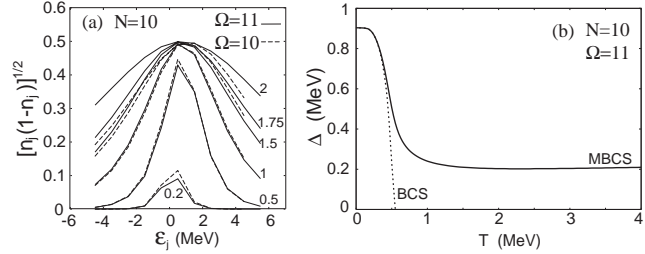}
\caption{\label{dNj}
(a) MBCS quasiparticle-number fluctuations 
$\delta{\cal N}_{j}$ within the PFM versus single-particle 
energies at several $T$. 
Lines connect discrete values to guide the eyes; 
numbers at the lines show the values of $T$ in MeV; (b) BCS and MBCS 
gaps for $N=$ 10 and $\Omega=$ 11 ($G=$ 0.4 MeV).}
\end{figure}

\begin{figure}                                                             
\includegraphics[width=20pc]{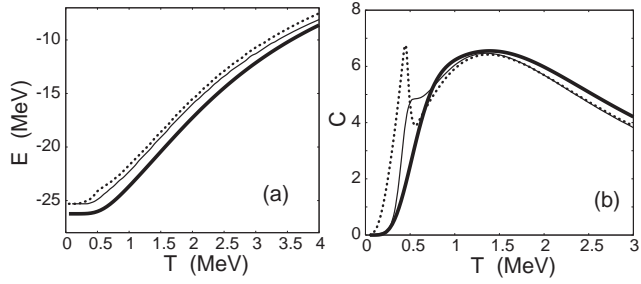}
\caption{\label{Etest}
Total energies (a) and heat capacities (b) within the PFM for ($N=$ 
10, $\Omega=$ 11, $G=$ 0.4 MeV) versus $T$. Dotted, thin-, 
and thick-solid lines denote the BCS, MBCS and exact results, 
respectively. A quantity equivalent to the self-energy term  
$-G\sum_{j}v_{j}^{4}$, not included within BCS and MBCS, 
has been subtracted from the exact total energy.}
\end{figure}

2) The virtue of the PFM is that it can be solved 
exactly in principle at $T=$ 0. However, 
at $T\neq$ 0 the exact solutions of a system with 
pure pairing do not represent a fully thermalized system. As a 
result, temperatures defined in different 
ways do not agree~\cite{Zele}. The limitation of the configuration space
with $\Omega=$ 10 causes a decrease of the heat capacity $C$ at $T_{\rm 
M}>$ 1.2 MeV(Schottky anomaly)~\cite{MHFB} (See Fig. 4 (c) of 
\cite{test}). 
Therefore, the region of $T>$ 1.2 MeV, generally speaking, is thermodynamically 
unphysical. The most crucial point here, however, 
is that such limited space deteriorates the criterion of 
applicability of the MBCS (See Sec. IV. A. 1 of ~\cite{MHFB}), which 
in fact requires that the line shapes of the 
quasiparticle-number fluctuations $\delta{\cal 
N}_{j}\equiv\sqrt{n_{j}(1-n_{j})}$ should be included symmetrically 
related to the Fermi level [Fig. 1 (f) of \cite{MHFB} is a good 
example]. 
The dashed lines in Fig. \ref{dNj} (a) 
shows that, for $N=$ 10 particles and $\Omega=$ 10 levels ($G=$ 
0.4 MeV), 
at $T$ close to 1.78 MeV, where the MBCS breaks down, $\delta{\cal 
N}_{j}$ are strongly asymmetric and large even for lowest and 
highest levels. 
At the same time, by just adding one more valence level ($\Omega=$ 
11) and keeping the same $N=$ 10 particles, we found that 
$\delta{\cal N}_{j}$ are rather symmetric related to the Fermi level 
up to much higher $T$ [solid lines in Fig. \ref{dNj} (a)]. This 
restores the balance in the summation of partial gaps 
$\delta\Delta_{j}$~\cite{MHFB}. 
As a result the obtained MBCS gap has no 
singularity at 0 $\leq T\leq$ 4 MeV [Fig. \ref{dNj} (b)]. 
The total energy and heat capacity obtained within the MBCS also 
agree better with the exact results than those given by the BCS 
[Fig. \ref{Etest}]. It is worth 
noticing that, even for such small $N$, adding one valence level increases the 
excitation energy $E^{*}$ by only $\sim$ 10$\%$ at $T=$ 2 MeV, while 
at $T<$ 2 MeV the values of $E^{*}$ for $\Omega=$ 10 and 11 are very 
close to each other.  
%%%%%%%%%%%%%%%%%%%%%%% Figure 1 %%%%%%%%%%%%%%%%%%%%%%%%%%%%%%
%%%%%%%%%%%%%%%%%%%%%%%%%%%%%%%%%%%%%%%%%%%%%%%%%%%%%%%%%%%%%%%%
%%%%%%%%%%%%%%%%%%%%%%% Figure 2 %%%%%%%%%%%%%%%%%%%%%%%%%%%%%%
%%%%%%%%%%%%%%%%%%%%%%%%%%%%%%%%%%%%%%%%%%%%%%%%%%%%%%%%%%%%%%%%
We also carried out the calculations for larger 
particle numbers $N$. This eventually increases 
$T_{\rm M}$, and also makes the line shapes of $\delta{\cal N}_{j}$ 
very symmetric at much higher $T$. 
For $\Omega=$ 50 and 100, e.g., we found $T_{\rm M}>$ 5 MeV, and 
the MBCS gap has qualitatively the same behavior as that of the solid line in Fig. \ref{dNj} (b) 
up to $T\sim$ 5 - 6 MeV. However, for large $N$ the exact solutions 
of PFM turn out to be impractical as a testing tool for $T\neq$ 0. 
Since all the exact eigenstates must be included in the partition function $Z$, 
and, since for $N=$ 50 e.g., the number of zero-seniority states alone 
already reaches $10^{14}$, the calculation of exact $Z$ 
becomes practically impossible.

\begin{figure}                                                             
\includegraphics[width=20pc]{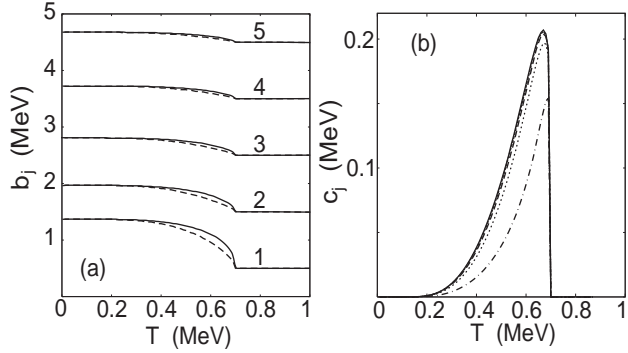}
\caption{\label{bjcj}
$b_{j}$ (a) and $c_{j}$ (b), 
obtained within BCS for 5 lowest levels in the PFM with 
$\Omega=$ 10 versus $T$. In (a) the solid and dashed 
lines represent $b_{j}$ and quasiparticle energies $E_{j}$, respectively.
In (b) the solid, dashed, dotted, and dash-dotted lines correspond
to levels 1 -- 5 in (a), respectively.}
\end{figure}

3) The principle of compensation of dangerous 
diagrams was postulated
to define the coefficients $u_{j}$ and $v_{j}$ 
of the Bogoliubov canonical transformation. 
This postulation and the variational calculation of 
$\partial{H'}/\partial{v_{j}}$ lead to Eq. (19) in  
\cite{test} for the BCS at $T=$ 0. It is justified so long as divergences can
be removed from the perturbation expansion of the ground-state energy.
However, at $T\neq$ 0 a $T$-dependent ground state does not 
exist. Instead, one should use the expectation values over the canonical or grand-canonical 
ensemble~\cite{MBCS2,MHFB}. 
Therefore, Eq. (19) of \cite{test} no longer holds at 
$T\neq$ 0 since the BCS gap 
is now defined by Eq. (7) of \cite{test}, instead of Eq. 
(3). Fig. \ref{bjcj} clearly shows how $b_{j}\neq E_{j}$ and $c_{j}\neq$ 
0 at $T\neq$ 0. 
This invalidates the critics based on Eq. (19) of \cite{test}.
%%%%%%%%%%%%%%%%%%%%%%% Figure 3 %%%%%%%%%%%%%%%%%%%%%%%%%%%%%%
%%%%%%%%%%%%%%%%%%%%%%%%%%%%%

In conclusion, the test of \cite{test}
is inadequate to judge the MBCS applicability 
because its results were obtained 
either in the $T$ region, where
the use of zero-$T$ spectra is no longer valid (for 
$^{120}$Sn and Ni), 
or within too limited configuration spaces (the PFM for $N=\Omega=$ 10 
or 2 major shells for Ni). Our 
calculations with a $T$-dependent spectrum for $^{120}$Sn, and 
within extended configuration spaces presented here show that the MBCS is a 
good approximation up to high $T$ even for a system with $N=$ 10 
particles.

We thank A. Volya for assistance in the exact solutions of the PFM.


\begin{thebibliography}{99}
\bibitem[1]{MBCS1}N. Dinh Dang and V. Zelevinsky, Phys. Rev. C {\bf 
64}, 064319 (2001).
\bibitem[2]{MBCS2}N. Dinh Dang and A. Arima, Phys. Rev. C {\bf 67}, 014304 (2003).
\bibitem[3]{MHFB}N.D. Dang and A. Arima, Phys. Rev. C {\bf 68}, 
014318 (2003).
\bibitem[4]{Moretto}L.G. Moretto, Phys. Lett. B {\bf 40}, 1 (1972).
\bibitem[5]{Egido}J.L. Egido, P. Ring, S. Iwasaki, and H.J. Mang, Phys. 
Lett. B {\bf 154}, 1 (1985).
\bibitem[6]{exp}T. Tsuboi and T. Suzuki, J. Phys. Soc. Jap. {\bf 42}, 
437 (1977); K. Kaneko and M. Hasegawa, Phys. Rev. C {\bf 72}, 
024307 (2005).
\bibitem[7]{test}V.Yu. Ponomarev and A.I. Vdovin, Phys. Rev. C {\bf 72}, 
034309 (2005).
\bibitem[8]{note}$\langle H\rangle_{\rm MBCS}$ is analytically equal 
to ${\cal E}_{\rm MBCS}$ by definition 
($\sim Gv_{j}^{4}$ neglected). 
Regarding footnote [11] of \cite{test}, no heat capacity was reported in \cite{MBCS2}.
\bibitem[9]{Bonche}P. Bonche, S. Levit, and D. Vautherin, Nucl. Phys. 
A {\bf 427}, 278 (1984).
\bibitem[10]{Zele}V. Zelevinsky and A. Volya, Phys. Part. Nucl. {\bf 
66}, 1829 (2003).
\end{thebibliography}
\end{document}